\documentclass{article}

\usepackage{arxiv}

\usepackage[utf8]{inputenc} 
\usepackage[T1]{fontenc}    
\usepackage{hyperref}       
\usepackage{url}            
\usepackage{booktabs}       
\usepackage{amsfonts}       
\usepackage{nicefrac}       
\usepackage{microtype}      
\usepackage{lipsum}		
\usepackage{graphicx}
\usepackage{natbib}
\usepackage{doi}

\usepackage{enumitem}
\usepackage{overpic}
\usepackage{xcolor}
\usepackage{listings}
\lstdefinestyle{yaml}{
     basicstyle=\color{black}\footnotesize,
     keywordstyle=\color{black}\bfseries,
     string=[s]{'}{'},
     stringstyle=\color{blue},
     comment=[l]{:},
     commentstyle=\color{black},
     morecomment=[l]{-},
     numbers=left,
     numbersep= 0pt,
     stepnumber=1,
     xleftmargin=0.2cm,
     numberstyle=\color{black},
     breaklines=true
 }

\title{A Flexible Architecture for Web-based GIS Applications using Docker and Graph Databases}

\author{ \href{https://orcid.org/0000-0003-2259-1254}{\includegraphics[scale=0.06]{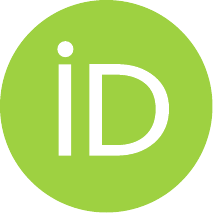}\hspace{1mm}Yves Annanias} \\
	Image and Signal Processing Group\\
	Leipzig University\\
	Leipzig, Germany\\
	\texttt{annanias@informatik.uni-leipzig.de} \\
	\And
	\href{https://orcid.org/0000-0002-5356-6732}{\includegraphics[scale=0.06]{orcid.pdf}\hspace{1mm}Daniel Wiegreffe} \\
	Image and Signal Processing Group\\
	Leipzig University\\
	Leipzig, Germany\\
	\texttt{daniel@informatik.uni-leipzig.de} \\
}

\renewcommand{\shorttitle}{Web-based GIS Architecture}

\hypersetup{
pdftitle={A Flexible Architecture for Web-based GIS Applications using Docker and Graph Databases},
pdfauthor={Yves Annanias, Daniel Wiegreffe},
}

\begin{document}
\maketitle

\begin{abstract}
Regional planning processes and associated redevelopment projects can be complex due to the vast amount of diverse data involved. 
However, all of this data shares a common geographical reference, especially in the renaturation of former open-cast mining areas.
 
To ensure safety, it is crucial to maintain a comprehensive overview of the interrelated data and draw accurate conclusions.
This requires special tools and can be a very time-consuming process.
A geographical information system (GIS) is well-suited for this purpose, but 
even a GIS has limitations when dealing with multiple data types and sources. 
Additional tools are often necessary to process and view all the data, which can complicate the planning process.
Our paper describes a system architecture that addresses the aforementioned issues and provides a simple, yet flexible tool for these activities.
The architecture is based on microservices using Docker and is divided into a backend and a frontend.
The backend simplifies and generalizes the integration of different data types, while a graph database is used to link relevant data and reveal potential new relationships between them.
Finally, a modern web frontend displays the data and relationships. 
\end{abstract}

\section{Introduction}
Geographical planning processes are complex, cover multiple domains and require large amounts of data that need to be interlinked to enable planning assessments and activities.
Especially in the context of urban planning, regional structural change, or renaturation of former opencast mining areas, the amount of data to be considered is large and diverse. 
In addition to expert opinions and legal requirements, other relevant data includes geographical information such as on-site sensor measurements, weather data, and 3D laser scans (LIDAR data).
The data required for this is usually multi-modal and often has to be prepared extensively before it can be used by a GIS, assuming the system offers import and visualization for this at all.
There are also some public sources that can be queried directly, although it is important to know how long the data is kept at this source. 
For legal reasons, all information contributing to a decision often have to be retained for a specified time period. 
Public sources may also have restrictions on the amount and frequency of data downloads to minimize server load.

After importing the individual data sets into the GIS, they are not yet fully linked with each other. 
Although it is possible to make statements about their geographical location and draw corresponding conclusions, this still requires a significant amount of manual work. 
For example, text passages must be linked to the corresponding areas on the map and examined for their applicability using current sensor data.
Finally, for an existing system, there is still the issue of how it can be expanded and how corresponding updates can be distributed to all users.

To address the aforementioned issues, we developed a comprehensive GIS.
In previous papers, we have described some detailed aspects of this system and demonstrated how various data sets can be visualized in combination. 
Describing the architecture on which this system is based is our contribution in this paper.
For that, requirements that individual parts of the system must fulfill are summarized.  
Altogether, we provide a description of an overall system that can 
    react flexibly to changing conditions, 
    handle large amounts of data, 
    expand flexibly, and
    with which relationships can be drawn between otherwise separate data sets.
Finally, we address some learnings we have gained.

Our project partner is the Lausitzer und Mitteldeutsche Bergbau-Verwaltungsgesellschaft mbH (\href{https://www.lmbv.de/}{LMBV}), which is responsible to renaturalize former opencast mines in the areas of Lausitz and Central Germany and prepare them for subsequent use. 
We received an extensive test data set from them, which includes geographical data, project data, and expert reports. 
This data set was supplemented and enriched with various public data sets, including weather and environmental data.

\section{Related Work}
GIS has numerous potential applications, particularly in environmental tasks and urban planning, which rely on the analysis of spatio-temporal and georeferenced data. 
A GIS can aid in the interpretation of this data, whereby visualization plays a major role \cite{Andrienko2003, Goodchild:2007}.
Container virtualization based on Docker is another technology with numerous applications, especially in cloud services where efficient application deployment is crucial \cite{docker_review}.
We will briefly cover both technologies below.\\\\
\textbf{Geodata and GIS}
GIS tools provide assistance in decision-making, for example during flood disasters \cite{li2017visual}, where hydrological parameters may be relevant \cite{schlegel2013determining}.
Climate change can also increase the likelihood of disasters, and GIS allows for the linking of simulation results with other factors to investigate their effects \cite{lei2015interactive}.
Other challenges arise from spatial ecology data for ecosystem restoration measures \cite{eligehausen2013blendgis} in addition to renaturation.
Another area of application is the health sector \cite{Healthcare}, especially in the case of pandemics such as Covid-19, where instructions for action can be derived from the analysis of the data \cite{covid19}.
Also to facilitate the energy transition in relation to climate change, it is important to consider the use of renewable energies. 
Therefore, it is necessary to provide an overview of energy consumption \cite{kolditz2014webgl} and carbon emissions \cite{energy_carbon}.
For sustainable urban development, GIS offers the opportunity to plan solar and wind parks \cite{solar_wind}.
This often poses new challenges, as legal and political restrictions exclude certain areas, where these mostly textual information thus concern geographical areas and must be related to them \cite{energy_restrictions}.
In all the applications listed, however, the main task is always to collect and link different, sometimes unstructured data, such as text data and spatial data, and to store them for seamless tracking. 
The challenges include correlating and aggregating the various data types \cite{covid19}.
However, the combination of different tools (e.g., GIS and BIM) offers new possibilities in the development of architectural and environmental units \cite{BIM_GIS}.
\\\\
\textbf{Microservices and Docker}
Docker is ideal for developing applications in isolation. Each container includes all necessary components, simplifying development, distribution, and deployment. 
Microservices, in particular, can then take full advantage of optimized technologies by encapsulating them from one another \cite{docker}.
Efficient use of server resources is also crucial for optimal user experience in web applications. 
Services on the server should scale automatically based on the load, and there are already frameworks available to assist with this \cite{docloud}.
In conclusion, combining Docker and GIS is a well-known concept.
For example, when dealing with big data, it is often necessary to store the data in a location other than the client's system. 
This may involve distributing the data across multiple systems and performing extensive analyses in a suitable format \cite{gis_bigdata}.
The example of a power grid GIS demonstrates that scalable and efficient applications can be developed with reduced effort \cite{gis_docker}.
Other frameworks also exist that combine container-based applications and GIS to provide efficient solutions \cite{gis_container_cloud}.
We utilize these techniques to manage a wide variety of data types that are integrated into a GIS. 
Our particular focus is on linking these data together and allowing for flexible extension and growth of the system.

\begin{figure*}[tb]
	\centering
	\includegraphics[width=0.975\linewidth]{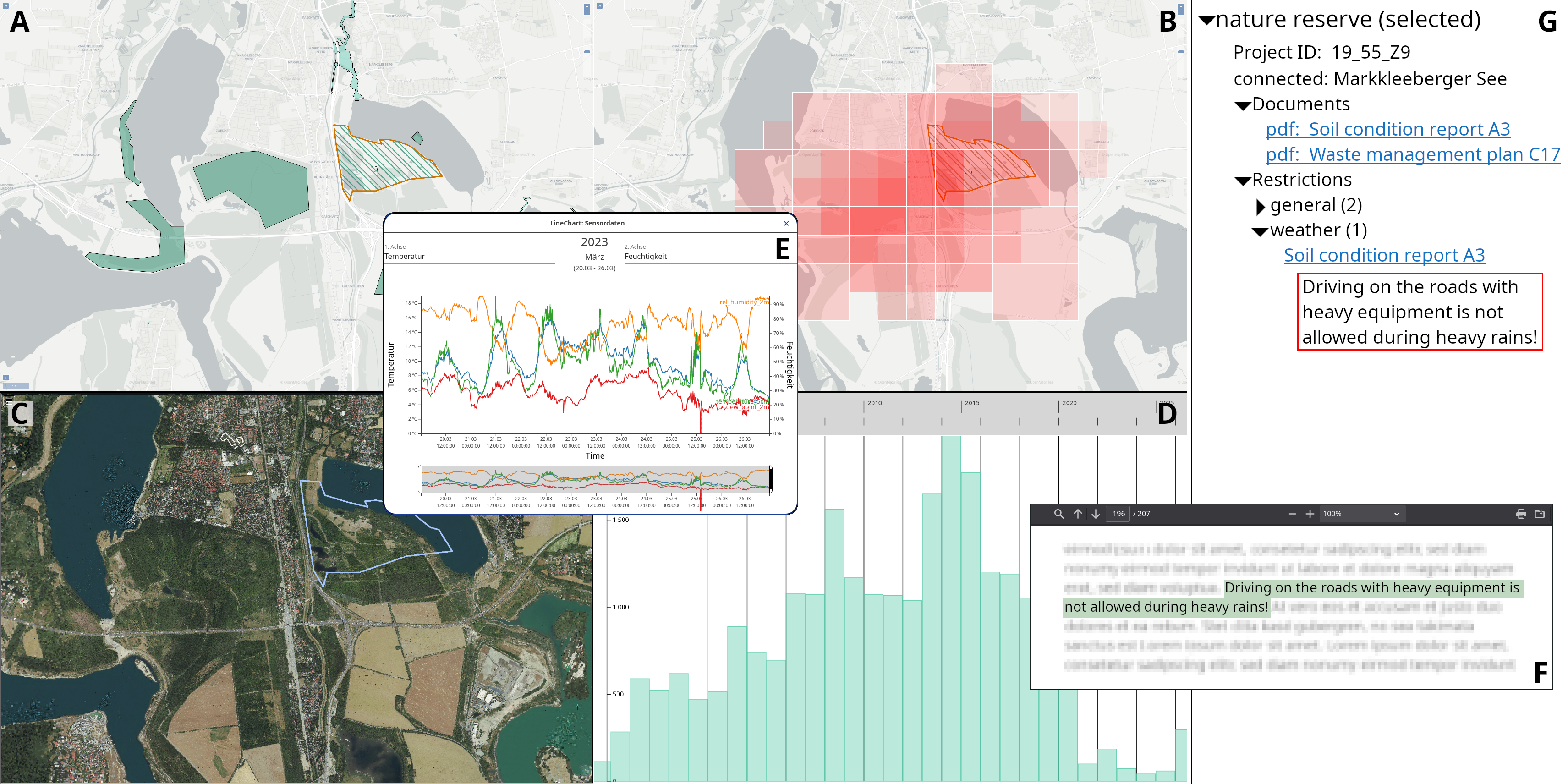}
	\caption{\label{fig:frontend}
		 Interlinked maps and information panels of the frontend. 
		 A) Projects with valid weather restrictions. 
		 B) Different feature categories are aggregated into a heatmap visualization. 
		 C) LIDAR data for the region.
		 D) Projects with time restrictions grouped by month.
		 E) Measured environmental data.
		 F) PDF Viewer with marked sentence.
		 G) Text data indicating a current weather restriction. 
	}
\end{figure*}

\section{Functionalities of the System}
Before the architecture of the system is described in more detail, a brief overview of its functionalities is given here (Figure \ref{fig:frontend}). 
Additional and more detailed descriptions of the visualization and interaction capabilities can be found in the referenced publications.

The GIS addresses the issue of visual clutter that arises when multiple polygons, including area and point data, are displayed simultaneously on a map and overlap either partially or completely \cite{YA_IDS}.
The polygons are grouped into different categories, whereby overlapping areas can reflect important information, but may not be recognizable. 
To address this, the system interactively aggregates different polygons and categories and displays them using a heatmap (Figure \ref{fig:frontend}B). 
Furthermore, the main map can be divided into multiple smaller, coordinated maps (Figure \ref{fig:frontend}A-C), allowing for an overview to be maintained despite the greater distance between data points.

Another data set comprises numerous text documents. 
These documents consist of expert reports on the topic of renaturalization, as well as reports from the active period of opencast mining. 
The documents themselves contain crucial and legally relevant statements, such as requirements to perform a specific operation or prohibitions on performing specific operations under certain conditions.
Locating these statements within the vast amount of documents manually is time-consuming, tedious and error-prone.
Using an active learning approach, the system automatically finds the relevant statements, classifies them and extracts them for further use \cite{schroder2021mining}.
The GIS can therefore not only list all documents, but also provides these extracted information (Figure \ref{fig:frontend}G).
The system always stores and references the original data as the automatic classification could be imperfect and sometimes more context is needed to understand an extracted legal information.
This allows a statement to be selected in the system, whereupon an integrated PDF viewer opens the document on the exact page from which the sentence was extracted.
The exact location on the page is also visually highlighted (Figure \ref{fig:frontend}F).

This also simplifies project management, where a project refers to various redevelopment tasks that are grouped together as a single unit in the workflow~\cite{borst22}.
Furthermore, projects can be grouped by additional topics based on the classification of the attached documents using the active learning approach.
This eliminates the timely need to review all documents in a project to check for statements on a particular topic.
With the grouping, a single click on a topic displays all projects (polygons) that contain such a document.
As a result of processing, the entire text corpus of all documents is now digital available. 
Any topics, statements or other information that have not yet been specifically analyzed and extracted can still be found using a simple full-text search.

By integrating weather data, such as from the German Weather Service (\href{https://www.dwd.de/EN/Home/home_node.html}{DWD}), the GIS can directly validate certain statements \cite{annanias2022interactive}. 
If threshold values defined in any of the documents are exceeded due to weather conditions, actions may need to be rescheduled. 
The GIS automatically recognizes valid statements and provides users with all necessary information to understand why a statement is valid (Figure \ref{fig:frontend}A and G), also the source of the statement in the document is indicated (Figure \ref{fig:frontend}F).
Other types of sensor data and environmental information, such as air quality, pollution levels, and water levels from the German Environment Agency (\href{https://www.umweltbundesamt.de/en}{UBA}), are also integrated (Figure \ref{fig:frontend}E).
There is already an established infrastructure for collecting and managing this data from various sources \cite{DBLP:conf/gi/WindolphWZF21}.
Time-dependent restrictions (e.g., breeding seasons) can also be extracted from the documents and visualized in a meaningful context in the GIS (Figure \ref{fig:frontend}D) \cite{EnvirVisTime}.

In addition, the system also provides the option to visualize other types of data, like LIDAR data, where the point data can be classified using categories such as buildings, vegetation, and roads \cite{GiLidar}. 
The tool Potree \cite{Potree}, an open-source web application, can be used to visually integrate this data into the system, because it is suitable for visualizing massive point clouds generated by LIDAR.
Combining Potree with the system allows the LIDAR data to be linked with other data in the system (Figure \ref{fig:frontend}C) using the coordinated maps layout.

\section{System Architecture}
The presented functionalities are only a part of the system and additional functionalities are still under development, which will make the system more complex. 
Data processing must be targeted, and some parts of the system must exchange data with each other.
Linking data is also crucial for in-depth analysis. 
To ensure efficient system functionality and keep complexity manageable, a system architecture is necessary, which we will describe in this section.
Please also refer to Figure 1 in the supplement for a system overview.

The entire system is divided into three main parts.
The first part, the frontend, deals exclusively with the display of the data and all visualizations and offers specific interactions. 
The second part, the backend takes care of data creation, storage and processing. 
On request from the frontend, it provides necessary data in an appropriate form and type and preprocesses the data (through filtering and aggregation) to reduce the amount of data that have to be transferred.
The third part, a graph database (as part of the backend) efficiently links different data with each other. 

With changing user requirements and the constant integration of new types of data, the entire system must be able to respond flexibly to these changes.
This places additional requirements on visualizing, processing, linking, and distributing the data.
Therefore, the following sections provide a detailed description of each part along with their respective requirements.



\subsection{Frontend}
The frontend takes care of the display of the map material, the visualization of the data, 
    the interactions and additional functionalities of a GIS 
    and must meet the following requirements:

\begin{itemize}[itemindent=0.45cm]
    \item [\textbf{RF1}]\label{req:RF1} The user interface should be easily accessible for a wide range of systems without significant technical requirements. 
    \item [\textbf{RF2}]\label{req:RF2} Updates to individual components or the backend should be rolled out quickly.
    \item [\textbf{RF3}]\label{req:RF3} It should be highly modular to enable new components for displaying data efficiently and without major dependencies.
    \item [\textbf{RF4}]\label{req:RF4} As many of the tools used in the project are web applications, these applications should be integrated directly into the frontend if possible in order to create a unified interface.
\end{itemize}

\paragraph*{Web Application}
The frontend was developed as a pure web application that can be executed in any modern web browser, which only needs a stable internet connection.
Apart from this, there are no other special system requirements.
Furthermore, a simple reload of the website is also sufficient to reflect all changes of the system, whereby \textbf{RF1} and \textbf{RF2} are fulfilled. 

By using the JavaScript framework \href{https://vuejs.org/}{Vue}, the frontend is therefore based purely on HTML, JavaScript and CSS.
Vue itself is based on the Model-View-ViewModel (MVVM) approach and, with its components, offers the possibility of implementing the code in a flexible, modular and, most importantly, reusable way. 
A component contains the encapsulated code for processing data as well as templates that are based on valid HTML and provide the UI for the component. 
A reactivity system automatically reflects changes to the data in the view. 
In addition to reusability, the components have the advantage that different visualizations can also be implemented separately from each other.
As a result, the entire frontend is designed as a modular system, as required by \textbf{RF3}.

\paragraph*{Integration of Web Applications} 
For some applications, suitable visualization tools already exist.
Implementing these from scratch is ineffective and involves a lot of effort. 
On the other hand, a separate tool with its own UI is then required for different tasks. 
If there are several applications, they have to be distributed manually on the screen or in different tabs in the browser, so switching between applications can lead to errors due to context change for the user.
Therefore, the combination of these applications offers advantages, but is often difficult or impossible to achieve, as many applications do not provide an API for this.
Data exchange can then be difficult, making integration beyond visual integration challenging. 
An example of how visual integration and data exchange could work for web applications is provided below.

The ability to integrate web applications using \textit{iFrames} (a standard HTML component) is a well known way to integrate web applications into other web applications.
This allows required applications to be opened and managed from the main application. 
The location of these applications can be configured, e.g., to appear as a new tab or as an additional visual component within the web application,
    giving the user the impression that both applications are one comprehensive interface. 
If the source code of the integrated application is available, as with Potree, it can be extended. 
This makes it possible for all the tools to exchange information with each other using \textit{postMessages} (a feature of JavaScript), which further strengthens the connection between these applications.
As a result, they can respond directly to events triggered by any application in the set.

Using \textit{iFrames} and \textit{postMessages} eliminates the need to download data from the first application, convert it if necessary, and upload it to the second application.
The data exchange happens automatically in the background without any further action by the user apart from a simple click.
The resulting direct connection of the applications also eliminates the need to download the same data from the server multiple times as it is already available, which reduces the load on the server.
An example is shown in Figure \ref{fig:frontend}.
Potree fits into the overall GIS view and can be arranged as an additional map (following the multiple connected maps layout from \cite{YA_IDS}). 
When a polygon is selected in map A, it is automatically forwarded to the Potree frontend (map C) where it is displayed alongside the LIDAR data.
This seamless integration allows for easy access to both applications without any additional effort for the user and fulfills \textbf{RF4}.

\subsection{Backend}
The backend manages the data and delivers it to the frontend in a suitably form.
For this, the following requirements must be met:

\begin{itemize}[itemindent=0.45cm]
    \item [\textbf{RB1}]\label{req:RB1} The backend should be modular and flexible. 
                                        A specific component should have only a few dependencies on other parts (e.g., only for data exchange).
                                        Then, details, such as frameworks and database systems can be easily exchanged without the need to adapt other components.
    \item [\textbf{RB2}]\label{req:RB2} Furthermore, a single part should be easy to set up, simplifying both the development phase and the final installation of the backend.
                                        This includes taking into account the potential problem of installing the correct packages without version conflicts.
    \item [\textbf{RB3}]\label{req:RB3} Individual components must be able to communicate with each other in order to exchange data. 
    \item [\textbf{RB4}]\label{req:RB4} The data should be accessible through a standardized interface, making it irrelevant for the user (and the frontend) to know the specific components or their implementation details, as complexity should be hidden.
    \item [\textbf{RB5}]\label{req:RB5} Only authorized users should have access to certain data.
\end{itemize}

\paragraph*{Docker}
Our solution is mainly based on a microservice architecture using \href{https://www.docker.com/}{Docker}.
This enables the creation of multiple dedicated containers for specific and largely independent tasks, 
    which are also isolated from the rest of the system.
Therefore, each task could define and install its own dependencies.
As a result, the system has a modular structure, as required by \textbf{RB1}.
  
\begin{figure}[b]
	\centering
	\includegraphics[width=1\linewidth]{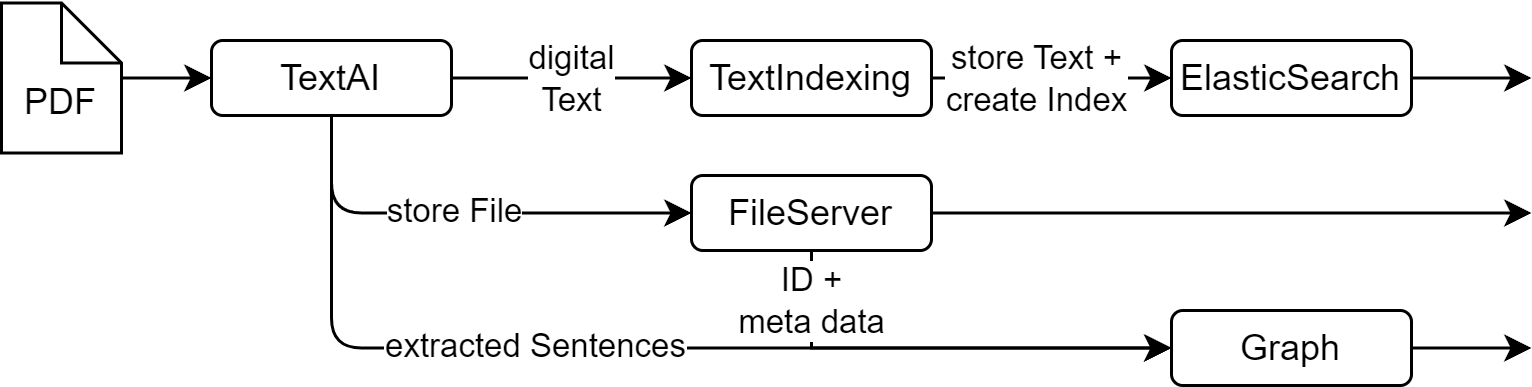}
	\caption[Component]{\label{fig:TextData}
		 Overview of the document processing using different containers for specific sub-tasks.
	}
\end{figure}

In the following, the use of such containers is illustrated by the example of the document processing, as shown in Figure \ref{fig:TextData}.
The \textit{TextAI} container utilizes an active learning and an optical character recognition approach to process documents, such as PDFs (for further information, see \cite{schroder2021mining}).
The document itself is then forwarded to a \textit{FileServer}, which stores all documents for later use and provides them on request. 
The \textit{TextAI} container then extracts the entire text and forwards it to the \textit{TextIndexing} container. 
In addition, specific sentences are also classified, extracted, and forwarded to a \textit{Graph} database for data mining purposes (see Section \ref{Database}).

Such containers are easily defined, as the example of the \textit{TextIndexing} container in Listing \ref{listing:indexer} shows.
This container reads the previously extracted text corpora, performs some preprocessing steps, and then passes the data over to an \textit{ElasticSearch} container that stores an index on the text data, enabling search queries to be performed.
For this, a file with the text data is passed in Line~\ref{listing:indexer_text}. 
The whole data is processed by a python script.
Therefore, the necessary packages are installed using the correct package versions (Lines \ref{listing:indexer_req} and \ref{listing:indexer_reqIns}).
Next, at Line \ref{listing:indexer_start}, the indexing is started.
The script processes the text data and then sends the data to an \textit{ElasticSearch} server running in its own container.
This modular and separated creation of individual containers with all their dependencies satisfies \textbf{RB2}.

\begin{lstlisting}[style=yaml, caption=Service: Text Indexing, label=listing:indexer, escapechar=|]
    FROM python:3.8-slim |\label{listing:indexer_text}|
    COPY TextCorpusData.csv TextCorpusData.csv
    COPY indexer.py indexer.py
    COPY requirements.txt requirements.txt |\label{listing:indexer_req}|
    RUN pip install -r requirements.txt |\label{listing:indexer_reqIns}|
    CMD python indexer.py --index_name main_index --csv_path TextCorpusData.csv |\label{listing:indexer_start}|
\end{lstlisting}

\paragraph*{Data Exchange}
The text indexing script must communicate with the \textit{ElasticSearch} container to exchange data and to fulfill \textbf{RB3}.
This can be easily solved using Docker-Compose, a tool that simplifies the creation and starting of multiple Docker containers.
Listing  \ref{listing:compose} shows a short excerpt. 
In this listing, individual Docker files are defined as services, such as the indexing script at Line \ref{listing:compose_indexer}.
For ElasticSearch, we use the standard image as in Line \ref{listing:compose_elasticsearch_image}.
By naming the services, the indexing script can communicate with the server using the name \textit{elasticsearch} in the python code instead of an IP address:
\textit{es = Elasticsearch(hosts=[{"host": 'elasticsearch', "port": 9200, 'scheme': 'http'}])}.
The port is defined in Line \ref{listing:compose_port} of the compose file, but since it is written as ":9200" instead of "9200:9200", this service is only accessible within the docker network, but not outside of it.
This is particularly necessary for both \textbf{RB4} and \textbf{RB5}.
For more detailed information on defining the entire network, please refer to the Docker-Compose \href{https://docs.docker.com/compose/networking/}{Documentation}.

\lstset{morekeywords={services, elasticsearch, indexer}}
\begin{lstlisting}[style=yaml, caption=Docker Compose - Services, label=listing:compose, escapechar=|]
    services:
        elasticsearch:
            image: elasticsearch:7.11.1 |\label{listing:compose_elasticsearch_image}|
            environment: 
                - network.bind_host=0.0.0.0
            ports:
                - ":9200" |\label{listing:compose_port}|
        indexer: |\label{listing:compose_indexer}|
            build: ./indexer/
            depends_on:
                - elasticsearch
\end{lstlisting}

\paragraph*{Main Server}
The individual services can exchange information with each other, but none of them delivers data to an external client.
Requests from a client are sent directly to a \textit{MainServer}, which is the only service that can communicate directly with clients.
This service is therefore (together with the graph database) a key service in the backend 
    and again a separate docker container that is added as a service to the docker compose file.

This simplifies requests, as they run through just one interface, as required by \textbf{RB4}.
Therefore, there is no need for several different endpoints, which can be requested externally. 
This is important, because each endpoint having its own special API would complicate requests.
Therefore, this service is capable of requesting the various required data from the individual services 
    while it can handle the different formats and file types.
The \textit{MainServer} then prepares the resulting data, i.e., filters for unnecessary details, and aggregates values together if requested or necessary.
This makes it also possible to verify which statements in the results are valid by incorporating additional information (see Use Case 2 in Section \ref{Database}).
Finally, this service returns all data to the frontend in a standardized format.

Additionally, this approach reduces the amount of data to be transferred as it can be filtered locally. 
It also reduces the workload for the client, as the frontend only has to process a small amount of data and can focus exclusively on visualization.
As a result, the hardware requirements for the frontend are lowered.

\paragraph*{External Services}
The system is flexible and new services can be added as containers. 
However, there may be cases where complete services already exist and need to be used.
It would then be inconvenient to have to set them up independently, copy all the data and integrate all of it into the system.

An example of this is the Internet of Things (IoT) system that collects, processes and provides sensor and weather data from the UBA and DWD services (see \cite{DBLP:conf/gi/WindolphWZF21} for more details).
Instead of copying this service and the entire infrastructure into our own system, we provide a service that only performs requests on the IoT system. 
It would also be possible to request the corresponding data using the frontend, but as a separate request in the backend (\textit{MainServer}), there are two main advantages.
First, the frontend still has to request only one location (\textbf{RB4}). 
Second, the \textit{MainServer} can utilize the IoT data to link, filter, and aggregate it with the other data.
For example, this information can be used to validate certain statements on other partial data records (see Use Case 2 in Section \ref{Database}) or to display regionally filtered values, as shown in Figure \ref{fig:frontend}E. 

Potree is an example of a standalone server that can display its data in its own UI. 
This is useful for experts who only require this functionality without the rest of the GIS.
However, the UI can also be integrated directly into our frontend, allowing Potree to interact directly with the rest of the GIS (Figure \ref{fig:frontend}C).
A Potree plugin enables authorized users (\textbf{RB5}) to query specific data from the backend system and link it with LIDAR data (\textbf{RB3}).

\paragraph*{Security}
Data and access security are crucial concerns, especially since our text documents contain personal data, trade secrets, and security information.
Therefore, it is essential to ensure the security of the data, which we have divided into two parts.
The first part covers security during the connection between the frontend and backend.
An encrypted https connection is ensured by a TLS~/~SSL certificate. 
As the \textit{MainServer} is the only externally accessible service, a certificate must be available exclusively for this service. 
Communication internally on the backend takes place locally, and data then only flows through this single service acting as a proxy.

The second part involves sharing data only with authorized users, which requires user administration and a rights management. 
In our system, this is achieved by logging onto the server with a username and password, which generates a token that identifies all the rights granted to that person.
Each time a request is made to the backend, the token must be sent, and the backend will then determine the data that can be provided. 
In this way, \textbf{RB5} is fulfilled.
However, implementing such a system can be complex.
It is recommended to include additional protective measures, such as limiting the lifespan of tokens and implementing mechanisms to address compromised tokens.
Anyway, there are online services that fully implement and offer this service, which can be used with less effort.

\subsection{Data Linkage}\label{Database}
Before our project, the data was only distributed as isolated records in the backend.
This meant that if a user was looking for a person involved in certain projects, they had to search through all documents using the available text data. 
Answering such a request is easy with the \textit{ElasticSearch} service and the index created on the text data. 
The \textit{MainServer} must then filter the documents according to the associated projects. 
For this, a project ID is read from the document, which is used to filter the relevant projects.

The whole process can also be used to search for specific companies (or any other search terms). 
This supports the exploration of relationships between search results. 
For example, it may be useful to identify all projects where a particular expert was responsible and a specific company was assigned renaturation tasks.
If the same company is hired for another project, the named person can also be consulted based on their experience with the quality of the service provided or can simply serve as a contact person.

In this case, two queries are run on the text data: one for the person and and one for the company. 
Both result sets must then be filtered for the projects and need to be merged to identify projects that contain both entities simultaneously.
This example refers to two specific entities that are known. 
In contrast, general queries and typical data mining queries, such as identifying people who frequently appear in the context of certain companies, can generate a large load on the server.
However, the loose coupling of the data complicates such analyses.
To solve this problem, it is necessary to link the data points in advance and store the relationships between them explicitly. 
Therefore, we defined the following requirements:

\begin{itemize}[itemindent=0.45cm]
    \item [\textbf{RD1}]\label{req:RD1} The database must be able to handle structural changes as the system allows for flexibility in available data, including the addition or removal of records, data types, or containers.
    \item [\textbf{RD2}]\label{req:RD2} The database's data structure should mirror the actual structure to facilitate relationship establishment inexperienced users.
\end{itemize}

\paragraph*{Graph Database}
As individual data points in our project are linked to other data points through various types of relationships, 
    it is more appropriate to use a graph model instead of a typical relational model. 
This is especially true as these relationships are just as relevant as the data itself.
A graph database has the advantage of incremental build-up, allowing for integration of new data records and types without major structural changes (\textbf{RD1}).
This is because the separation of data from its storage and management structure is less relevant.
Hence, there is no need to establish a rigid schema in advance that is highly tailored to the specific data and thus inflexible to modifications.
The data structure generated in this way also reflects the real scenario, without the need to transform or normalize the inherent data structure and thus corresponds to a natural modeling (\textbf{RD2}). 
Representing knowledge is thus simplified and remains intuitive.
Angles and Gutierrez \cite{GDM} also mention these points in their introduction to graph data management.

The following two use cases demonstrate the straightforward modeling that addresses the aforementioned problem and demonstrates the benefits and strengths of data linkage.
A \href{https://neo4j.com/}{Neo4j} database is used for this purpose, which can be queried using \href{https://neo4j.com/developer/cypher/}{Cypher}.

\paragraph*{Use Case 1}

\begin{figure}[tb]
	\centering
	\includegraphics[width=0.6\linewidth]{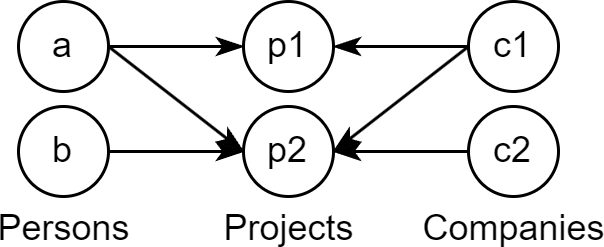}
	\caption[Component]{\label{fig:UseCase1}
		 Example graph showing how entities (persons / companies) extracted from the documents, are related to different projects.
	}
\end{figure}

By using graph modeling, the previous example of people and companies working together on a project can be implemented easily and efficiently.
First, a node is created for each project.
Instead of searching for the names of people or companies in the text data as before, these entities can be extracted beforehand using a Named Entity Recognition (NER).
Again, a node is created for each entity and linked with an edge to the corresponding project node, to which the documents belong.
This produces the graph depicted in Figure \ref{fig:UseCase1}.
Then only the following single Cypher query is required, to collect all projects (\textit{p2}) where person \textit{b} and company \textit{c2} appear together:\\
\textit{MATCH (p:Person \{name: "b"\})-[]→(project:Project)<-[]-(c:Company \{name: "c2"\}) RETURN project}\\
This eliminates the need to search the index of the text data multiple times.
The size of the text data or the size of the entire database in general is also barely relevant for a fast response.
Furthermore, the general query about the frequencies of any pairs can be obtained by just a minor addition, 
    which reveals that person \textit{a} and company \textit{c1} have already worked together on two projects (namely \textit{p1} and \textit{p2}):\\
\textit{MATCH (p:Person)-[]->(project)<-[]-(c:Company) RETURN p, c, COUNT(project)}

\paragraph*{Use Case 2} \label{usecaseWeather}
Another example is the system's ability to find areas that are subject to usage restrictions at a given time.
This involves searching, extracting and categorizing the text data for statements specifically about the weather (see \cite{annanias2022interactive} for more details).
Then, a new "Weather" node is created in the graph. 
Next, for each sentence that matches the topic, an edge is created pointing to the document from which the sentence originates. 
The extracted sentence and some meta data are noted on the edge itself. 
To check whether the described weather event is present, the graph is supplemented by nodes for sensors, measuring points, signaling systems, and so on (hereinafter referred to as sensor nodes).
An edge between such a sensor node and the area indicates that the measured values apply to the area (through overlapping or local proximity to the area). 

\begin{figure}[tb]
	\centering
	\includegraphics[width=0.7\linewidth]{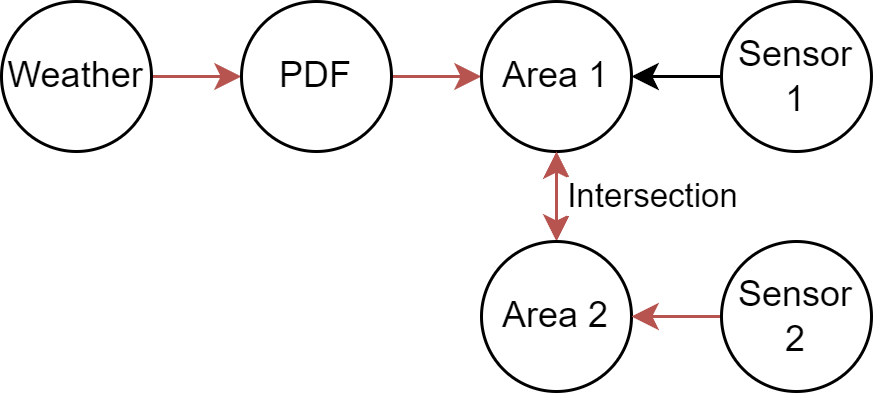}
	\caption[Component]{\label{fig:UseCase2}
		 Weather restrictions for an area can be checked using corresponding sensors nearby, even if the specified area does not have such a restriction directly (red path).
	}
\end{figure}

To check if there is a particular weather restriction for an area, it is sufficient to check whether there is a path from the weather node to a sensor node (see upper row in Figure \ref{fig:UseCase2}).
Then, the \textit{MainServer} can check the list of specified sensors and ask the IoT system if any of them exceed a threshold value.
By analyzing the overlap of the areas, statements can also be made about partial areas that do not have an explicit restriction mentioned in a attached document.
In this way, information can also be generated from potentially incomplete data (red path in Figure \ref{fig:UseCase2}), 
    because it is possible that the report was not prepared for the second area or was not stored correctly. 
However, as both areas partly cover the same region, it can be assumed that the same restrictions also apply there.
Enriched with additional information, the system can provide users with detailed explanations for active restrictions (Figure \ref{fig:frontend}G), identify their sources (Figure \ref{fig:frontend}F) and show all affected areas (Figure \ref{fig:frontend}A).
Circumstances that previously required manual checking with significant effort can now be quickly and automatically queried.
Furthermore, this use case is not limited to weather data, other time-dependent issues can also be analyzed using a similar approach \cite{EnvirVisTime}.

\section{Lessons Learned}
Despite the advantages of the presented system, we have also encountered problems occasionally, which we would like to mention briefly.

The project successfully utilized microservices with Docker, allowing for quick and flexible setup of individual services. 
Thus, this approach proved to be productive.
Additionally, different developers were able to independently solve subtasks.
However, maintaining a system that utilizes various paradigms, languages, and styles can be challenging. 
To mitigate this, it is essential to establish boundaries, such as limiting the selection of languages and paradigms, and providing comprehensive documentation. 

Additionally, different components need to exchange data and interact with each other and we have shown that communication between individual services can be established easily.
However, exchanging data can also be problematic. 
For instance, if a service is modified, the format of the data may change, making it incompatible with the next service unless it is adapted too. 
Standards are necessary to define the input and output formats of individual services and if a change is necessary, at least one other service must also be adapted.

Besides that, we have had positive experiences with the development and use of a web application. 
It was well received by our experts, stakeholders, and partners due to its simplicity of access via a single URL 
    and compatibility with almost any mobile device (depending on the size and resolution of the display).
Web development promotes platform-independent development and distribution of applications. 
However, it is important to note that browsers are also platforms with their own dependencies.
As a result, standards may be implemented differently by different browsers, such as Chrome and Firefox.
In our project, large quantities of simple character strings were initially transferred and processed differently, 
    resulting in significant performance differences between the two browsers.
Furthermore, default configurations vary between browsers, including how many resources a web application is allowed to use and the available functionalities. 
It is important to consider this, as it is usually unreasonable to expect users to adjust their browser settings.

Next, we used Potree to demonstrate its integration into our application. 
This integration offers clear advantages for users, making it easier to work with various tools, reducing errors, and complementing the strengths of the applications.
But not all applications could be easily integrated, and modifying the source code is often necessary. 
This reduces maintainability, as changes to the original system may cause customizations to stop working. 
A plugin system that ensures specific functionalities would improve stability. 
However, it would also be beneficial if applications offered interfaces for data exchange or interactions. 
Unfortunately, these are not standard and are only provided by a few applications.

\section{Future Work \& Conclusion}
Currently, the system only stores backups of the data.
Nevertheless, a comprehensive backup system is necessary not only for data but also for the recovery of subsystems in the event of failure to ensure smooth operations.
\href{https://docs.docker.com/engine/swarm/swarm-tutorial/}{Docker Swarm} can provide resilience by offering multiple replicas for containers and enabling load balancing. 
Alternatively, \href{https://kubernetes.io/}{Kubernetes} can also distribute services to different servers based on the generated load.
In addition, integrating further data sources would be beneficial.
Combining solar data and sunshine duration with LIDAR data can provide information on optimal locations for solar systems. 
Water level measurements, linked to weather data (i.e., rainfall), could be used for hazard defense and to issue fully automatic warnings.

All together, we have presented a system architecture for a real-world application that is specifically designed for the renaturation of former open-cast mining areas. 
The application can also be extended to other areas, and additional components can be easily connected to the Docker network to provide their data.
The separation of backend and frontend has several advantages. 
It ensures that data processing and storage are independent of the display.
As a result, the frontend requires minimal resources and can operate on standard mobile systems with a modern browser and internet connection. 
Updates are automatically rolled out by reloading the application, resulting in cost savings, especially in a corporate context.
Finally, the development process included several meetings with LMBV experts and other stakeholders who believe in its usefulness.
It is worth mentioning that they quickly understood the data structure modeling using a graph. 
They associated it with mind maps and train timetables, which made the modeling more vivid.

\section*{Acknowledgment}
This research was supported by the Development Bank of Saxony (SAB) under Grant 100400221.

\bibliographystyle{unsrtnat}
\bibliography{references}  






\end{document}